\documentclass[10pt,aps,twocolumn]{revtex4}
\usepackage{graphicx}
\usepackage{amsmath}
\usepackage{setspace}

\begin{document}

\title{Further developments in correlator product states: deterministic optimization and energy evaluation}
\author{Eric Neuscamman$\mathrm{{}^\ast}$, Hitesh Changlani$\mathrm{{}^\dag}$, Jesse Kinder$\mathrm{{}^\ast}$, and Garnet Kin-Lic Chan$\mathrm{{}^\ast}$}
\affiliation{$\mathrm{{}^\ast}$Department of Chemistry and Chemical Biology, Cornell University, Ithaca, New York 14853\\
             $\mathrm{{}^\dag}$Department of Physics, Cornell University, Ithaca, New York 14853}
\date{August 29, 2010}

\begin{abstract}
Correlator product states (CPS) are a class of tensor network wavefunctions applicable to strongly correlated problems in arbitrary dimensions.
Here, we present a method for optimizing and evaluating the energy of the CPS wavefunction that is non-variational but entirely deterministic.
The fundamental assumption underlying our technique is that the CPS wavefunction is an exact eigenstate of the Hamiltonian, allowing the energy
to be obtained approximately through a projection of the Schr\"odinger equation.
The validity of this approximation is tested on two dimensional lattices for the spin-$\frac{1}{2}$ antiferromagnetic Heisenberg
model, the spinless Hubbard model, and the full Hubbard model.
In each of these models, the projected method reproduces the variational CPS energy to within 1\%.
For fermionic systems, we also demonstrate the incorporation of a Slater determinant reference into the ansatz, which allows CPS to act as a
generalization of the Jastrow-Slater wavefunction.
\end{abstract}

\maketitle

\section{Introduction}
\label{sec:introduction}

Tensor network wavefunctions represent an efficient approach to modeling strongly interacting quantum systems.
While the density matrix renormalization group (DMRG) algorithm \cite{WHITE:1992:dmrg,SCHOLLWOCK:2005:dmrg} and
corresponding matrix product state (MPS) have been extremely successful in one dimensional systems, the application
of tensor networks to two dimensions remains a work in progress.
The pair entangled product state (PEPS) wavefunction
\cite{CIRAC:2004:peps,CIRAC:2007:peps,CIRAC:2008:ipeps,WEN:2008:peps,XIANG:2008:trg,XIANG:2009:trg,YANG:2009:peps,
      VIDAL:2009:ipeps,VIDAL:2010:fpeps,CIRAC:2010:fpeps,VERSTRAETE:2010:fpeps}
represents the natural extension of the MPS in two dimensions,
but lacks the tractability with respect to evaluating expectation values that the MPS enjoys in one dimension.
Similarly, the multi-scale entanglement renormalization ansatz (MERA) \cite{VIDAL:2007:mera,VIDAL:2008:mera}
has been limited by computational cost.
Recently, a simpler class of tensor network wavefunction has been proposed by both some of the present authors
\cite{CHANGLANI:2009:cps} and others
\cite{MEZZACAPO:2009:entangled_plaquettes,MEZZACAPO:2010:entangled_plaquettes,VERSTRAETE:2010:cgtn}.
This wavefunction, known as the correlator product state (CPS), the entangled plaquette state (EPS), or the complete graph
tensor network, seeks to correlate neighboring lattice sites directly, without the use of auxiliary bond indices.
It is closely related to the string bond state, \cite{CIRAC:2008:string_bond,CIRAC:2010:string_bond} although the
correlators are simple tensors and do not involve any underlying MPS structure.
As discussed previously \cite{CHANGLANI:2009:cps}, the CPS ansatz also encompasses a number of other wavefunctions,
including the Huse-Elser ansatz \cite{HUSE:1988:huse_elser}, the Laughlin wavefunction
\cite{LAUGHLIN:1983:frac_quantum_hall}, and the toric code \cite{KITAEV:2003:toric_code}.
So far in its development, the CPS ansatz has been studied through the use of variational Monte Carlo (VMC) techniques,
which provide efficient procedures for its optimization and the evaluation of expectation values in two and higher
dimensions \cite{CHANGLANI:2009:cps,MEZZACAPO:2009:entangled_plaquettes,MEZZACAPO:2010:entangled_plaquettes}.

Here we present an alternative approach to optimizing and evaluating the energy of the CPS wavefunction that is entirely
deterministic, i.e. no stochastic sampling is involved.
Much like the coupled cluster (CC) formalism commonly used in quantum chemistry \cite{BARTLETT:2007:cc_review}, our
method relies on assuming that the CPS wavefunction is  an exact eigenstate of the Hamiltonian.
Under this assumption, the Schr\"{o}dinger equation may be projected against a carefully chosen bra state in order
to yield an approximate, non-variational expression for the system's energy that can be evaluated without stochastic
sampling.
To be accurate, this projected energy functional requires an optimization of the CPS wavefunction that differs from
that used in VMC. As the accuracy of the functional depends on how close the CPS state is to a true eigenstate, the
wavefunction is optimized by satisfying a set of projected Schr\"{o}dinger equations, which comprise a subset of the
conditions necessary for a wavefunction to be a Hamiltonian eigenstate.
Central to our approach is the formulation of the CPS wavefunction as a product of local, invertible operators acting
on a reference wavefunction.
In previous work, the CPS reference has been taken as an equally weighted sum of the states in the Hilbert space,
which we name the uniform reference.
Here, we also consider a Slater determinant reference for fermionic systems, which allows CPS to act as a generalization
of the Jastrow-Slater wavefunction long used in variational Monte Carlo simulations
\cite{JASTROW:1955:jastrow_slater,FOULKES:2001:QMC_Review}.

To evaluate the accuracy of our projected energy functional, we have applied both the VMC CPS and the projected CPS (PCPS)
methods to three model Hamiltonians on two-dimensional lattices.
Our first test case is the spin-$\frac{1}{2}$ antiferromagnetic Heisenberg model, which is currently the most
widely studied model for the CPS ansatz. \cite{CHANGLANI:2009:cps,MEZZACAPO:2009:entangled_plaquettes}
We then tested our method on two fermionic models, the spinless and regular Hubbard models, in which we employed the
Slater determinant reference.
In all cases, we find that the PCPS method reproduces the results of VMC CPS to within 1\% in the ground state energy.
This represents a first proof of principle that the technique of projecting the Schr\"{o}dinger equation can be effectively
applied to a tensor network wavefunction.
 
We begin by reviewing the structure of the CPS ansatz and presenting its formulation in terms of operators
acting on a reference wavefunction (Sec.\ \ref{sec:ansatz}).
We then review how the energy may be evaluated stochastically through VMC (Sec.\ \ref{sec:vmc_energy}) before presenting
the projected energy functional (Sec.\ \ref{sec:projected_energy}) and how a system of projected Schr\"{o}dinger equations may be used
to optimize the CPS ansatz (Sec.\ \ref{sec:optimization}).
Results are then presented for the Heisenberg model (Sec.\ \ref{sec:heisenberg}), spinless Hubbard model (Sec.\ \ref{sec:spinless_hubbard}), and full
Hubbard model (Sec.\ \ref{sec:hubbard}).
Finally, we summarize the main points and discuss future directions for CPS wavefunction research (Sec.\ \ref{sec:conclusions}).

\section{Theory}
\label{sec:theory}

\subsection{The CPS Ansatz}
\label{sec:ansatz}

Central to the definition of the CPS wavefunction is the concept of a correlator, which 
directly encodes correlations between lattice sites.
Consider a subset of the lattice sites, $P=\{i_1, i_2 \ldots i_l\}$, where each site $i$ carries quantum states $|n_i\rangle$ such as spin
or fermionic degrees of freedom. A correlator  with domain  $P$ is defined by its amplitudes 
$C_P^{n_{i_1} n_{i_2} \ldots n_{i_l}}$  for each configuration in the Hilbert space of $P$. The total CPS wavefunction $|\Psi\rangle$ is 
built from multiple correlators on separate (possibly overlapping) sets of the sites. In a simple  CPS, for a  lattice configuration
$|n_1 n_2 \ldots n_k\rangle$, the corresponding wavefunction amplitude $\langle n_1 n_2 \ldots n_k|\Psi\rangle$ is given by the
product of  correlator amplitudes, i.e.
\begin{align}
\label{eqn:occ_num_wfn}
|\Psi \rangle = \sum_{n_1 n_2 ... n_k}  \prod_{P} C_P^{n_{i_1} n_{i_2} \ldots n_{i_l}}  | n_1 n_2 ... n_k \rangle.
\end{align}
For example, for  nearest neighbor correlators in one-dimension with open boundary conditions, Eq. (\ref{eqn:occ_num_wfn})  is 

\begin{align}
|\Psi \rangle = \sum_{n_1 n_2 ... n_k} {C}^{n_1 n_2} C^{n_2 n_3} \ldots C^{n_{k-1} n_k}  | n_1 n_2 ... n_k \rangle.
\end{align}
The CPS wavefunction can also be written in a more general way so that the correlators appear as operators acting on a
reference state.
We first define the projection operator $\hat{\rho}_{n_i} = |n_i\rangle\langle n_i|$ which projects to the subspace in
which site $i$ is in state $n_i$.
For example,
\begin{align}
\label{eqn:site_proj_op}
\hat{\rho}_{n_i^\prime} | n_1  \ldots n_i \ldots n_k \rangle = \delta^{n_i^\prime}_{n_i} | n_1  \ldots n_i \ldots n_k \rangle,
\end{align}
where $\delta$ is the Kronecker delta.
The correlator operator for domain $P = \{i_1, i_2, ..., i_l\}$ can now be defined by combining the correlator amplitudes
with a product of projection operators for each site in $P$,
\begin{align}
\label{eqn:corr_op}
\hat{C}_P \equiv \sum_{n_{i_1} n_{i_2} ... n_{i_l}} C^{n_{i_1} n_{i_2} ... n_{i_l}}_P \hat{\rho}_{n_{i_1}} \hat{\rho}_{n_{i_2}} \ldots \hat{\rho}_{n_{i_l}}.
\end{align}
We then write the CPS wavefunction as a product of correlator operators acting on a reference state $|\Phi\rangle$,
\begin{align}
\label{eqn:op_wfn}
|\Psi\rangle =  \prod_{P } \hat{C}_P  |\Phi\rangle.
\end{align}
In order for this wavefunction to be equivalent to that in Eq.\ (\ref{eqn:occ_num_wfn}), we must choose $|\Phi\rangle$ to be
the sum of all possible lattice configurations with equal coefficients, a state we refer to as the uniform reference, 
\begin{align}
|\Phi\rangle = \sum_{n_1 n_2 \ldots n_k} |n_1 n_2 \ldots n_k\rangle. \label{eqn:reference}
\end{align}
In this work, we usually restrict the summation in Eq. (\ref{eqn:reference}) to states with particular quantum numbers.
For example, for a spin system, we sum only over those spin configurations with a given $S_z$ value, while for a
fermionic system, we sum only over configurations with a given $N$ and given $S_z$.

One can also use other reference functions for $|\Phi\rangle$, in which case the correlator operators can be
seen as providing some additional correlation beyond what is present in the reference.
In practice, only some reference functions will be useful.
Specifically, they must not interfere with the efficient evaluation of expectation values.
In this work, we will also explore the Slater determinant as a reference function, which allows the CPS wavefunction to act
as a generalization of the Jastrow-Slater wavefunction \cite{JASTROW:1955:jastrow_slater,FOULKES:2001:QMC_Review}.

\subsection{The Monte Carlo Variational Energy}
\label{sec:vmc_energy}

In previous work, \cite{MEZZACAPO:2009:entangled_plaquettes,MEZZACAPO:2010:entangled_plaquettes,CHANGLANI:2009:cps}
the CPS energy has been evaluated by means of Monte Carlo (MC) sampling.
For a general wavefunction,
\begin{align}
\label{eqn:mc_wfn}
|\Psi\rangle = \sum_{n_1 n_2 ... n_k} \Psi^{n_1 n_2 ... n_k} | n_1 n_2 ... n_k \rangle,
\end{align}
the energy may be written as
\begin{align}
\label{eqn:mc_energy}
        E = \frac{\langle\Psi|H|\Psi\rangle}{\langle\Psi|\Psi\rangle}
          = & \sum_{n_1 n_2 ... n_k} \frac{|\Psi^{n_1 n_2 ... n_k}|^2}{\langle\Psi|\Psi\rangle} E_{\mathrm{L}}(n_1 n_2 ... n_k),
\end{align}
where the local energy $E_{\mathrm{L}}(n_1 n_2 ... n_k)$ is defined by
\begin{align}
\nonumber
& E_{\mathrm{L}}(n_1 n_2 ... n_k) = \\
& \ \ \ \ \sum_{n_1^\prime n_2^\prime ... n_k^\prime} \frac{\Psi^{n_1^\prime n_2^\prime ... n_k^\prime}}{\Psi^{n_1 n_2 ... n_k}}
          \langle n_1 n_2 ... n_k | H | n_1^\prime n_2^\prime ... n_k^\prime \rangle.
\label{eqn:mc_local_energy}
\end{align}
A Markov chain can then be used to sample the probability distribution $|\Psi^{n_1 n_2 ... n_k}|^2/\langle \Psi|\Psi\rangle$ and compute the
overall energy as the average of the sampled local energies.
As discussed in previous reports \cite{CHANGLANI:2009:cps,MEZZACAPO:2009:entangled_plaquettes,CIRAC:2008:string_bond},
the CPS wavefunction is well suited to MC sampling because the amplitudes $\Psi^{n_1 n_2 ... n_k}$ are easy to evaluate.
When $|\Phi\rangle$ is chosen as the uniform reference, the amplitudes are simply a product of
correlator values, as in Eq.\ (\ref{eqn:occ_num_wfn}).
When $|\Phi\rangle$ is allowed to take on other forms, the contribution of the reference function, the factor $\langle n_1 n_2 ... n_k |\Phi\rangle$, must be
included as well.
In order to evaluate the energy by MC sampling in practice, we are therefore limited to forms of $|\Phi\rangle$ for
which this factor can be evaluated efficiently. Two forms of $|\Phi\rangle$ have long been used in quantum Monte Carlo simulations.
For the Slater determinant, $\langle n_1 n_2 ... n_k |\Phi\rangle$ is equal to a determinant of orbital coefficients
that can be evaluated in at most $O(N_p^3)$ time by an LU decomposition, where $N_p$ is the number of particles in the system.
In practice, this cost can be reduced to $O(N_p^2)$ during a Markov chain iteration by using the matrix determinant lemma and
the Sherman-Morrison formula \cite{SHERMAN:1950:sherman_morrison,CEPERLEY:1977:many_fermion_MC}. 
Another  reference function for which sampling is efficient is the BCS wavefunction \cite{SCHRIEFFER:1957:bcs} and its number
projected form, the antisymmetrized geminal power \cite{POPLE:1953:agp,ANDERSON:1987:RVB,SORELLA:2003:Jastrow_AGP_Atoms}.
However, we will not explore the BCS reference here.

\subsection{The Deterministic Projected Energy Functional}
\label{sec:projected_energy}

\begin{figure}[t]
\centering
\includegraphics[width=8.5cm,angle=0]{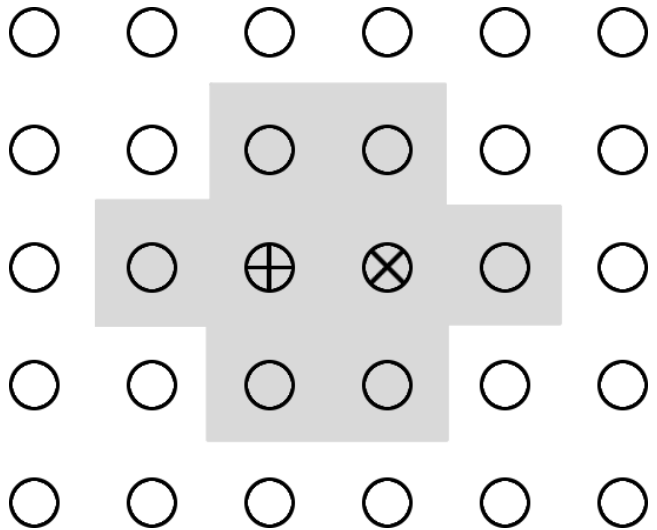}
\caption{An example 2D lattice showing the cluster (shaded) resulting from nearest neighbor pair correlators
         around a hopping operator.  The site touched by the creation operator is marked with a ``+''
         and the site touched by the destruction operator by an ``x''.  Computing the projected CPS
         energy requires summing over only the configurations of the different clusters produced
         by the Hamiltonian's operators, rather than over all lattice configurations.
         The energy for a local Hamiltonian can thus be evaluated in $O(N d^M)$ time, where $N$ is
         the size of the lattice, $d$ the number of configurations of a site, and $M$ the number
         of sites in a cluster.}
\label{fig:cluster}
\end{figure}

In the current work, in addition to using a Monte Carlo evaluation of the energy, we also explore
another method  which does not require the use of stochastic sampling.
We first make an approximation in which we assume that the CPS
wavefunction is  an eigenstate of the Hamiltonian,
\begin{align}
\label{eqn:assume_eigenstate}
H |\Psi\rangle & \simeq E |\Psi\rangle \Rightarrow H  \prod_{P } \hat{C}_P  |\Phi\rangle \simeq E  \prod_{P } \hat{C}_P  |\Phi\rangle.
\end{align}
We then define an inverse correlator product bra state $\langle \tilde{\Psi}|$, which is obtained by
 left multiplying $\langle\Phi|$ with a product of inverse correlator operators
\begin{align}
\label{eqn:inverse_bra}
\langle \tilde{\Psi}| = \langle \Phi |\prod_P \hat{C}_P^{-1}
\end{align}
where
\begin{align}
\label{eqn:inv_corr_op}
\hat{C}^{-1}_P = \sum_{n_{i_1} n_{i_2} ... n_{i_l}}  \frac{1}{C^{n_{i_1} n_{i_2} ... n_{i_l}}_P} \hat{\rho}_{n_{i_1}} \hat{\rho}_{n_{i_2}} \ldots 
\hat{\rho}_{n_{i_l}}.
\end{align}
This gives an approximate energy functional that is exact if $|\Psi\rangle$ is an eigenstate,
\begin{align}
\label{eqn:proj_energy}
E = \langle \tilde{\Psi} | H | \Psi\rangle.
\end{align}
Provided that the correlators are not too large or long-ranged, Eq.\ (\ref{eqn:proj_energy}) can be evaluated efficiently
by considering the different operators inside the Hamiltonian individually.
Consider the energy contribution $E_{x y}$ from a hopping operator $a^{\dag}_x a_y$,
\begin{align}
\label{eqn:hop_energy}
 E_{x y} \simeq \langle \tilde{\Psi} | a^\dag_x a_y | \Psi\rangle = \langle\Phi|  \prod_{Q} \hat{C}^{-1}_Q a^{\dag}_x a_y \prod_{P} \hat{C}_P  |\Phi\rangle.
\end{align}
This hopping operator is associated with two sites $x$ and $y$.
We can divide correlators (and their inverse correlators) into two kinds: those which touch
the sites $x$ or $y$ and those which do not. Correlators which do not touch $x$ or $y$ can be commuted
past the the hopping operator  and thus cancel with their inverse partners.
The set of  correlators which touch $x$ or $y$ define a connected cluster of sites,
$\Omega_{xy}=\{c_1, c_2, \ldots, c_{xy}\}$ (see Fig.\ \ref{fig:cluster} for an example).
The central point is that, as long as the domain size of the correlators is independent of lattice size,
the size of the cluster $\Omega_{xy}$ is also independent of lattice size, and this allows expectation values
of the form (\ref{eqn:hop_energy}) to be evaluated efficiently (i.e. with a cost scaling polynomially in the lattice size).

Commuting terms past $a^\dag_x a_y$, the product of correlators in (\ref{eqn:hop_energy}) becomes 
\begin{align}
\label{eqn:hop_energy_simplified}
E_{x y} = \langle\Phi|  \prod_{Q \in S_{x y}} \hat{C}^{-1}_Q \ a^{\dag}_x a_y \ \prod_{P \in S_{x y}}\hat{C}_P  |\Phi\rangle,
\end{align}
where $S_{x y}$ denotes the set of correlators that touch $x$ or $y$.
For a more explicit expression, we separate the correlators into their amplitude and projection operator components.
The expectation value of the projection operators define a many-body reduced density-matrix (RDM) $\gamma$,
\begin{align}
\nonumber
  \gamma^{n_{c_1}  \ldots n_{c_{xy}}}_{n_{c_1}^\prime \ldots n_{c_{xy}}^\prime} = 
  \langle \Phi| & \hat{\rho}_{n_{c_1}}  \ldots \hat{\rho}_{n_x} \hat{\rho}_{n_y} \ldots \hat{\rho}_{n_{c_{xy}}} \\
& \ \ a^\dag_x a_y \hat{\rho}_{n_{c_1}^\prime}  \ldots \hat{\rho}_{n_x^\prime} \hat{\rho}_{n_y^\prime} \ldots \hat{\rho}_{n_{c_{xy}}^\prime} |\Phi\rangle.
\label{eqn:density_matrix}
\end{align}
Combining this density matrix with the correlator amplitudes, we obtain
\begin{align}
\label{eqn:density_matrix_energy}
E_{xy} = \mathop{\sum_{n_{c_1} \ldots n_{c_{xy}}}}_{n_{c_1}^\prime \ldots n_{c_{xy}}^\prime} \gamma^{n_{c_1}  \ldots n_{c_{xy}}}_{n_{c_1}^\prime  \ldots n_{c_{xy}}^\prime} \prod_{P \in S_{x y}} \frac{C_P^{n_{i_1}^\prime \ldots n_{i_l}^\prime}}{C_P^{n_{i_1} \ldots n_{i_l}}}.
\end{align}
For each term in the sum, the quotient of correlators is easily evaluated.
However, we must also evaluate the relevant RDM element, which limits us to references for which $\gamma$ is readily available.
The uniform reference's RDM elements are trivial to evaluate (even with particle number and $S_z$ restrictions) 
while those of a Slater determinant can be computed as a determinant of one-body RDM elements in $N^3$ time,
where $N$ is the number of lattice sites.
With the RDM elements in hand, the final task is to sum over all terms in Eq.\ (\ref{eqn:density_matrix_energy}).
Due to the sparsity of $\gamma$, there are at most $d^M$ non-zero terms, where $d$ is the size of the single-site Hilbert space
and $M$ is the number of sites in the cluster.
For sufficiently small and local correlators (i.e. for small enough clusters), the summation can be carried out
exactly without resorting to importance sampling.
To summarize, each operator within the Hamiltonian defines a cluster, on which we compute a sum of correlator ratios
and RDM elements.
The total energy is then the sum of each operators' contribution.

Although the energy functional of Eq.\ (\ref{eqn:proj_energy}) allows for a non-stochastic evaluation of the CPS wavefunction's
energy, it possesses a number of shortcomings that should be discussed.
First, it is not a variational form, as the initial assumption that the wavefunction is an exact Hamiltonian
eigenfunction is only an approximation.
Second, the functional is only tractable for small, local correlators.
For example, if all lattice site pairs were used to define correlators, then each hopping operator would create a cluster
the size of the lattice, and the functional would be no more tractable than the variational energy expression.
Similarly, local correlators that contain a large number of sites may lead to clusters whose configurations are too
numerous to sum over explicitly.
A final concern with the energy functional is the form of the Hamiltonian.
In lattice models, one rarely encounters nonlocal operators or operators that touch more than two sites, so the
clusters formed around them will be manageable provided the correlators are small and local.
However, in quantum chemistry, the Hamiltonian contains the long range Coulomb interaction, which manifests as a nonlocal
four-site operator. This would severely restrict the size of correlator for which summation over the cluster would be feasible.

\begin{center}
\begin{table*}[t]
\caption{Energies per site of the antiferromagnetic spin-$\frac{1}{2}$ Heisenberg model with
         periodic boundary conditions and local correlators.
         The correlator sizes are given with the method names.
         Energies are in units of $J$, with the number in parentheses representing the uncertainty
         in the final digit.
         See Sec.\ \ref{sec:heisenberg} for details.}
\begin{tabular}{  c   r@{.}l   r@{.}l   r@{.}l   r@{.}l   r@{.}l  }
\hline\hline
                        Lattice Size
 & \multicolumn{2}{ c }{\hspace{2mm} PCPS 2x2    \hspace{1mm} }
 & \multicolumn{2}{ c }{\hspace{1mm} VMC CPS 2x2 \hspace{1mm} }
 & \multicolumn{2}{ c }{\hspace{1mm} VMC CPS 3x3 \hspace{1mm} }
 & \multicolumn{2}{ c }{\hspace{1mm} VMC EPS 4x4$\mathrm{{}^a}$ \hspace{1mm} }
 & \multicolumn{2}{ c }{\hspace{1mm} SSE$\mathrm{{}^b}$ \hspace{1mm} } \\
\hline
     4x4        & \hspace{3mm} -0&694226 \hspace{1mm} & \hspace{5mm} -0&69432(1) \hspace{1mm} & \hspace{5mm} -0&70150(1) \hspace{1mm} & \hspace{5mm} -0&7016(1) \hspace{1mm} & \hspace{1mm} -0&701777(7) \hspace{1mm} \\
     6x6        & \hspace{3mm} -0&668615 \hspace{1mm} & \hspace{5mm} -0&66806(2) \hspace{1mm} & \hspace{5mm} -0&67615(1) \hspace{1mm} & \hspace{5mm} -0&6785(2) \hspace{1mm} & \hspace{1mm} -0&678873(4) \hspace{1mm} \\
     8x8        & \hspace{3mm} -0&665920 \hspace{1mm} & \hspace{5mm} -0&66165(1) \hspace{1mm} & \hspace{5mm} -0&66989(1) \hspace{1mm} & \hspace{5mm} -0&6724(3) \hspace{1mm} & \hspace{1mm} -0&673487(4) \hspace{1mm} \\
    10x10       & \hspace{3mm} -0&664910 \hspace{1mm} & \hspace{5mm} -0&65900(1) \hspace{1mm} & \hspace{5mm} -0&66776(1) \hspace{1mm} & \hspace{5mm} -0&6699(3) \hspace{1mm} & \hspace{1mm} -0&671549(4) \hspace{1mm} \\
\hline\hline
\multicolumn{11}{l}{$\mathrm{{}^a}$ Entangled plaquette state, see Ref.\ \cite{MEZZACAPO:2009:entangled_plaquettes}.} \\
\multicolumn{11}{l}{$\mathrm{{}^b}$ Stochastic series expansion, see Ref.\ \cite{SANDVIK:1997:heisenberg_sse}.} \\
\end{tabular}
\label{tab:heisenberg_results}
\end{table*}
\end{center}

\subsection{Optimizing the Wavefunction by Projection}
\label{sec:optimization}

When using the variational Monte Carlo energy functional, we can obtain the ground state
by minimizing the energy. Several different algorithms can be used. In Ref.\ \cite{CHANGLANI:2009:cps}, we
used a ``sweep'' algorithm where a single correlator is updated at a time by solving an effective Schr\"odinger equation.
For the variational Monte Carlo results reported in this work, we have employed a steepest descent optimization scheme
that updates all correlators simultaneously.

The deterministic projected energy functional, however, is not a variational expression. Consequently we
require a different method to optimize the correlators in the CPS when using this functional.
Indeed, accurate results from the projected functional
require correlators that make the wavefunction as close as possible to a Hamiltonian eigenstate.
This optimization condition is slightly different from minimizing the energy, and will in general produce slightly different correlators.

We begin as in the previous section by assuming that the CPS wavefunction is a true Hamiltonian
eigenstate and therefore satisfies the Schr\"{o}dinger equation,
\begin{align}
\label{eqn:eigenstate_condition}
(H - E)  \prod_{P } \hat{C}_P  |\Phi\rangle \simeq 0.
\end{align}
We then seek to enforce this condition approximately by applying a number of projections.
We define a set of bra states $\langle \tilde{\Psi}_P^{n_{i_1} n_{i_2} \ldots n_{i_l}}|$ obtained
from the inverse CPS bra state in Eq. (\ref{eqn:inverse_bra})
\begin{align}
\langle \tilde{\Psi}_P^{n_{i_1} n_{i_2} \ldots n_{i_l}}|  = \langle \tilde{\Psi}| \hat{\rho}_{n_{i_1}} \hat{\rho}_{n_{i_2}} \ldots \hat{\rho}_{n_{i_l}} \ : \ i_1, i_2 \ldots i_l \in P,
\end{align}
where the projectors are those associated with a given correlator $C_P$. Projecting with these bra states, we obtain
\begin{align}
\label{eqn:proj_eigenstate_condition}
\Lambda^{n_{i_1} n_{i_2} ... n_{i_l}}_P = \langle \tilde{\Psi}_P^{n_{i_1} n_{i_2} \ldots n_{i_l}}| H - E | \Psi \rangle = 0.
\end{align}
By requiring that each $\Lambda^{n_{i_1} n_{i_2} ... n_{i_l}}_P$ vanish, we create a system of nonlinear
projected Schr\"{o}dinger equations that can be solved for the correlator amplitudes.
As with the energy, we can evaluate the contributions of the individual Hamiltonian elements to these equations
separately.
The local cluster formed around each Hamiltonian operator is the same as in the energy evaluation, except that the
projection operators $\hat{\rho}_{n_i}$ fix some of the sites' occupations.

In order to solve the system of equations given in Eq.\ (\ref{eqn:proj_eigenstate_condition}), we have taken two approaches.
First, the derivatives of each $\Lambda^{n_{i_1} n_{i_2} ... n_{i_l}}_P$ with respect to each correlator amplitude can be
evaluated, producing a Jacobian matrix that can be used in a standard Newton-Raphson procedure.
Alternatively, we may take the approach of constructing and diagonalizing a local Hamiltonian for each correlator.
This second approach arises from fixing all variables except for one correlator's amplitudes, which makes the equations
linear in those amplitudes.
The exact solution to this simplified system of linear equations is then equivalent to the lowest eigenvector of the
local Hamiltonian.
The local Hamiltonian for a correlator on sites $P = \{i_1, i_2 ... i_l\}$ is defined by
\begin{align}
\label{eqn:local_hamiltonian}
H^{n_{i_1} ... n_{i_l}}_{n_{i_1}^\prime  ... n_{i_l}^\prime} =
\langle\Phi| & \prod_{Q \neq P} \hat{C}^{-1}_Q \hat{\rho}_{n_{i_1}} \ldots \hat{\rho}_{n_{i_l}} \\
& \ \ \ (H - E) \hat{\rho}_{n_{i_1}^\prime} \ldots \hat{\rho}_{n_{i_l}^\prime} \prod_{Q^\prime \neq P} \hat{C}_{Q^\prime} |\Phi\rangle.
\nonumber
\end{align}
As before, we may evaluate this expression efficiently if we consider each Hamiltonian operator's contribution separately,
in which case we need only sum over the configurations of the local cluster around each operator.
Once we have evaluated the local Hamiltonian and found its lowest eigenvalue and eigenvector, we update the correlator
corresponding to $P$ with the eigenvector's elements.
This process is performed for each correlator, and then is repeated iteratively until $\Lambda$ is sufficiently small.

\section{Results}
\label{sec:results}

\subsection{Antiferromagnetic Heisenberg Model}
\label{sec:heisenberg}

We have optimized the CPS ansatz using the  variational Monte Carlo and  projected energy functionals, for the antiferromagnetic
spin-$\frac{1}{2}$ Heisenberg model on a number of periodic square lattices of edge length L.
The Hamiltonian, in which the parameter $J$ is assumed to be positive, is written as
\begin{align}
\label{eqn:heisenberg_hamiltonian}
H = J \sum_{\left \langle i, j \right \rangle} \vec{S}_i \cdot \vec{S}_j,
\end{align}
where the notation $\left \langle i, j \right \rangle$ indicates the set of all nearest neighbor pairs.
The limitations of the cluster size for the projected method (see Sec.\ \ref{sec:projected_energy}) restricted us to using 4-site (2x2) correlators.
The variational Monte Carlo method can practically support up to 16 site (4x4) correlators, results for which were reported earlier by
Mezzacapo et al \cite{MEZZACAPO:2009:entangled_plaquettes}.
We have reproduced the 4x4 results and also optimized the variational wavefunction for 4 site (2x2) and
9 site (3x3) correlators.
The results are summarized in Table \ref{tab:heisenberg_results} for $L$ = 4, 6, 8, and 10.
For 2x2 correlators, both the variational and projected methods produce energies with a relative error between
1 and 2\% of the essentially exact stochastic series expansion (SSE) \cite{SANDVIK:1997:heisenberg_sse}.
Thus we see that the eigenfunction assumption underpinning the projected method is already reasonable for 2x2 correlators.
We would naturally expect the projected and VMC methods to produce even closer energies to each other for larger correlators,
as in this case the wavefunction would be even closer to an exact Hamiltonian eigenstate.
By using larger correlators in the variational wavefunction, we see that the relative error can be systematically reduced.
The 3x3 correlators produce relative errors below 0.6\%, while the 4x4 correlators produce relative errors below 0.25\%.
In summary, the projected method produces reasonably accurate results on the 2D Heisenberg model that are comparable
to those of the variational method, however the limitation on correlator size prevents it from obtaining the higher
accuracies achieved with large correlators in the variational method.
\begin{figure}[t]
\includegraphics[width=8.5cm,bb= 84 58 285 251]{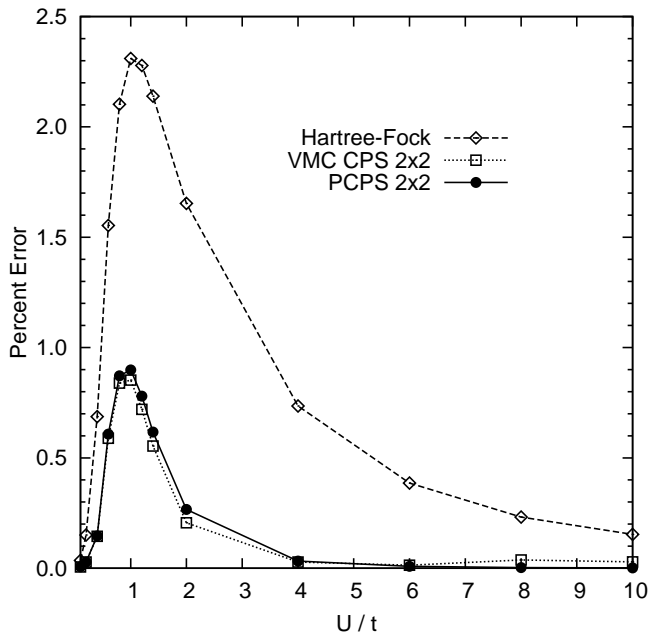}
\caption{Ground state energy errors for the 20 site (4x5) spinless Hubbard model with 10 particles and open boundary conditions.
         Exact results were computed using the ALPS program \cite{ALPS}.}
\label{fig:spinless_hubbard_4x5_n10}
\end{figure}
\begin{figure}[t]
\centering
\includegraphics[width=8.5cm,bb= 84 58 285 251]{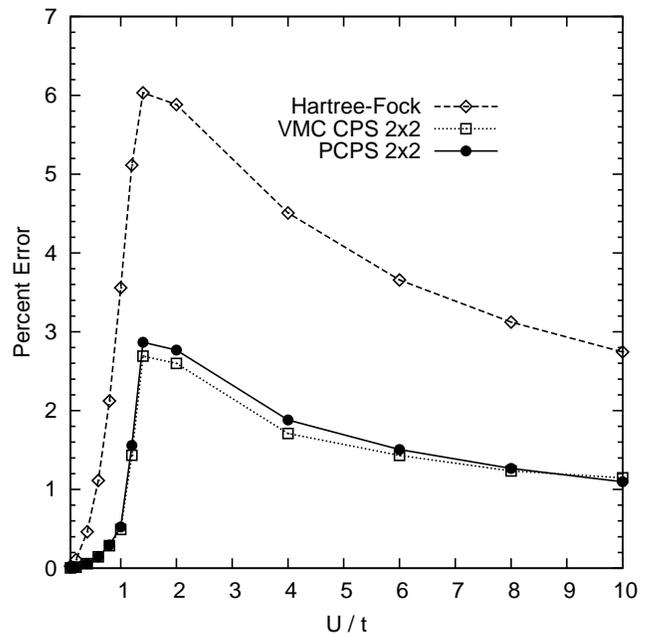}
\caption{Ground state energy errors for the 20 site (4x5) spinless Hubbard model with 9 particles and open boundary conditions.
         Exact results were computed using the ALPS program \cite{ALPS}.}
\label{fig:spinless_hubbard_4x5_n9}
\end{figure}

\subsection{Spinless Hubbard Model}
\label{sec:spinless_hubbard}

We have treated a 20-site (4x5) spinless Hubbard lattice with open boundary conditions using 4 site (2x2) correlators
and the Hartree-Fock Slater determinant reference.
While the CPS ansatz is capable of treating much larger lattices, we chose this lattice size in order to compare to exact results.
The Hamiltonian for the spinless Hubbard model is
\begin{align}
\label{eqn:spinless_hubbard_hamiltonian}
H = \sum_{\left \langle i, j \right \rangle} U a^\dag_i a_i a^\dag_j a_j - t ( a^\dag_i a_j + a^\dag_j a_i ),
\end{align}
in which $a^\dag_i$ and $a_i$ are the fermionic particle creation and destruction operators on site $i$,
and $\left \langle i, j \right \rangle$ represents the sum over all nearest neighbor site pairs.
The first term in Eq.\ (\ref{eqn:spinless_hubbard_hamiltonian}) represents a repulsive interaction between particles
on neighboring sites, while the second term represents hopping between sites.

The quality of the projected energy functional is sensitive to the quality of the reference function $|\Phi\rangle$.
Using the uniform reference is therefore undesirable, as it is clearly incorrect for large $U/t$.
We instead use a Slater determinant whose orbital coefficients have been variationally optimized through
Hartree-Fock theory.
Such a determinant satisfies the model exactly in the small $U/t$ limit for any filling, and in the large $U/t$ limit
at half filling.
Furthermore, we have observed that the determinant is also accurate for single hole doping at large $U/t$.

Results for half-filling and single hole doping are presented in Figs.\ \ref{fig:spinless_hubbard_4x5_n10} and
\ref{fig:spinless_hubbard_4x5_n9}, respectively.
For the case of half-filling, the CPS results using both the projected and variational functions 
are within 1\% of the exact energy for all values of the ratio $U/t$.
For single hole doping, the worst error is below 3\%.
If one considers the CPS wavefunction as a generalization of the Jastrow-Slater wavefunction, these accuracies are not extraordinary.
However, it should be stressed that the projected CPS results, which are essentially the same as the variational results,
require no stochastic sampling as is typically used in Jastrow-Slater simulations, but are obtained entirely deterministically.

\begin{center}
\begin{table*}[t]
\caption{Ground state energies for the Hubbard model at half filling with open boundary conditions.
         The DMRG results used m=1600 renormalized states.
         The correlator sizes used were 3-site for the 1D lattices and 4-site (2x2) for the 4x5 lattice.
         Energies are in units of $t$, with the uncertainty in the final digit placed in parentheses.
         See Sec.\ \ref{sec:hubbard} for details.}
\label{tab:hubbard_results}
\begin{tabular}{  c   r@{.}l   r@{.}l   r@{.}l   r@{.}l  }
\multicolumn{9}{ c }{$U/t=2$ \vphantom{$\frac{1}{\frac{1}{2}}$}} \\
\hline\hline
                        Lattice Size
 & \multicolumn{2}{ c }{\hspace{3mm} DMRG    \hspace{1mm} }
 & \multicolumn{2}{ c }{\hspace{3mm} RHF     \hspace{1mm} }
 & \multicolumn{2}{ c }{\hspace{3mm} PCPS    \hspace{1mm} }
 & \multicolumn{2}{ c }{\hspace{3mm} VMC CPS \hspace{1mm} } \\
\hline
     1x14        & \hspace{3mm} -11&279897 \hspace{1mm} & \hspace{5mm} -10&133544 \hspace{1mm} & \hspace{5mm} -11&241776 \hspace{1mm} & \hspace{5mm} -11&240(1) \hspace{1mm} \\
     1x18        & \hspace{3mm} -14&653987 \hspace{1mm} & \hspace{5mm} -13&219131 \hspace{1mm} & \hspace{5mm} -14&592961 \hspace{1mm} & \hspace{5mm} -14&591(1) \hspace{1mm} \\
     1x22        & \hspace{3mm} -18&029379 \hspace{1mm} & \hspace{5mm} -16&307287 \hspace{1mm} & \hspace{5mm} -17&946260 \hspace{1mm} & \hspace{5mm} -17&947(2) \hspace{1mm} \\
     4x5         & \hspace{3mm} -20&127521 \hspace{1mm} & \hspace{5mm} -18&800678 \hspace{1mm} & \hspace{5mm} -19&920320 \hspace{1mm} & \hspace{5mm} -19&917(1) \hspace{1mm} \\
\hline\hline
\multicolumn{9}{ c }{       } \\
\multicolumn{9}{ c }{$U/t=4$ \vphantom{$\frac{1}{\frac{1}{2}}$}} \\
\hline\hline
                        Lattice Size
 & \multicolumn{2}{ c }{\hspace{3mm} DMRG    \hspace{1mm} }
 & \multicolumn{2}{ c }{\hspace{3mm} RHF     \hspace{1mm} }
 & \multicolumn{2}{ c }{\hspace{3mm} PCPS    \hspace{1mm} }
 & \multicolumn{2}{ c }{\hspace{3mm} VMC CPS \hspace{1mm} } \\
\hline
     1x14        & \hspace{3mm}  -7&672349 \hspace{1mm} & \hspace{5mm}  -3&133544 \hspace{1mm} & \hspace{5mm}  -7&631100 \hspace{1mm} & \hspace{5mm}  -7&556(1) \hspace{1mm} \\
     1x18        & \hspace{3mm}  -9&965398 \hspace{1mm} & \hspace{5mm}  -4&219131 \hspace{1mm} & \hspace{5mm}  -9&842409 \hspace{1mm} & \hspace{5mm}  -9&770(3) \hspace{1mm} \\
     1x22        & \hspace{3mm} -12&259082 \hspace{1mm} & \hspace{5mm}  -5&307287 \hspace{1mm} & \hspace{5mm} -12&042344 \hspace{1mm} & \hspace{5mm} -11&964(4) \hspace{1mm} \\
     4x5         & \hspace{3mm} -14&404488 \hspace{1mm} & \hspace{5mm}  -8&800678 \hspace{1mm} & \hspace{5mm} -13&384297 \hspace{1mm} & \hspace{5mm} -13&350(1) \hspace{1mm} \\
\hline\hline
\end{tabular}
\end{table*}
\end{center}

\subsection{Full Hubbard Model}
\label{sec:hubbard}

We have applied the CPS ansatz to the Hubbard model with open boundary conditions and half filling in both one and two dimensions.
The Hubbard Hamiltonian is
\begin{align}
\label{eqn:hubbard_hamiltonian}
H =    U \sum_i a^\dag_{i \uparrow} a_{i \uparrow} a^\dag_{i \downarrow} a_{i \downarrow}
     - t \sum_{\left \langle i, j \right \rangle} \sum_{\sigma=\uparrow,\downarrow} ( a^\dag_{i \sigma} a_{j \sigma} + a^\dag_{j \sigma} a_{i \sigma} ),
\end{align}
in which $a^\dag_{i \uparrow(\downarrow)}$ and $a_{i \uparrow(\downarrow)}$ are the fermionic creation and destruction operators
for particles with spin $\uparrow$ $(\downarrow)$, and the notation $\left \langle i, j \right \rangle$ refers to the set of
all nearest neighbor site pairs.
We use as our reference function the restricted Hartree-Fock (RHF) Slater determinant.
While this reference is far from ideal as it is qualitatively incorrect in the large $U/t$ limit, it is sufficient for illustrating
the central point we seek to convey.
Better energies could be obtained with an unrestricted determinant or a particle number projected BCS reference
\cite{POPLE:1953:agp,SCHRIEFFER:1957:bcs}, but in the present discussion we limit ourselves to asking how the projected and variational CPS formulations compare, and for this
purpose the RHF determinant is sufficient as a reference function.

Results for the ratios $U/t=2$ and $U/t=4$ are presented in Table \ref{tab:hubbard_results}.
For the lower ratio, the RHF reference produces energies in error by 6-10\%, which are reduced to 1\% or less after the optimization
of either the projected or variational CPS wavefunction.
More interesting is the fact that the projected and variational CPS energies differ from each other by less than 0.02\% for both the
one and two dimensional lattices.
For the case of $U/t=4$, the RHF reference is qualitatively incorrect with relative errors as high as 60\%.
Despite this poor starting point, both variational and projected CPS reduce the error to 2\% and 7\% in one and two dimensions, respectively.
These accuracies are worse than in the $U/t=2$ case, but this is primarily due to the poor reference and could be improved.
Crucially, the projected method still effectively reproduces the variational CPS energy, the relative difference between them
never exceeding 1\%.

\section{Conclusions}
\label{sec:conclusions}

We have shown that by applying certain projections to the Schr\"{o}dinger equation, the correlator product state (CPS) wavefunction can be optimized and
its energy approximated without the need for stochastic sampling.
The energies produced by this projected CPS method differ by less than 1\% from the corresponding variational Monte Carlo (VMC) results in tests on
three types of two dimensional systems:  the spin-$\frac{1}{2}$ antiferromagnetic Heisenberg model, the spinless Hubbard model,
and the full Hubbard model.
While the projection procedure is currently limited to smaller correlators than the VMC approach, 
it is nonetheless encouraging that the projected Sch\"{o}dinger equation technique can be applied to a tensor network wavefunction,
and it is our hope that this technique will find use with other tensor network ansatze.

We have also demonstrated that the CPS wavefunction may be usefully separated into a product of correlator operators acting on some
reference function.
For fermions, we have showed that reasonable results can be obtained through the use of a Slater determinant reference, which makes
the CPS ansatz a generalization of the Jastrow-Slater wavefunction.
For the Hubbard model, we employed a restricted determinant, and thus we expect further improvements in accuracy could be achieved by
simply breaking the spin restriction.
Other candidates for use as CPS references include the BCS and AGP wavefunctions, which should further improve on the Slater determinant and
will be explored in future work.

\section{Acknowledgments}
\label{sec:acknowledgments}

This work was supported by the National Science Foundation through the NSF Center for Molecular Interfacing as well as Grant No.\ CHE-0645380
and CHE-1004603.

\bibliographystyle{aip}
\bibliography{pcps.bib}

\end{document}